\documentclass[aps,pra,floatfix,showpacs,preprint,onecolumn]{revtex4-2}

\usepackage{graphicx}
\usepackage{amsmath, amsfonts, amssymb, bm}

\usepackage{amstext}
\usepackage{mathrsfs}
\usepackage{color}

\begin{document}
\title{Unveiling the transverse formation length of nonlinear Compton scattering}
\author{A.~Di Piazza}
\email{dipiazza@mpi-hd.mpg.de}
\affiliation{Max Planck Institute for Nuclear Physics, Saupfercheckweg 1, D-69117 Heidelberg, Germany}

\begin{abstract}
The process of emission of electromagnetic radiation does not occur instantaneously but it is ``formed'' over a finite time known as radiation formation time. In the ultrarelativistic regime, the corresponding (longitudinal) formation length is given by the formation time times the speed of light and controls several features of radiation. Here, we elucidate the importance of the transverse formation length (TFL) by investigating nonlinear Compton scattering by an electron initially counterpropagating with respect to a \emph{flying focus} laser beam. The TFL is related to the transverse size of the radiation formation ``volume'' and, unlike the longitudinal formation length, has a quantum origin. Being the TFL typically of the order of the Compton wavelength, where any laser field can be assumed to be approximately uniform, related quantum interference effects have been ignored. However, we show analytically that if the focus in a flying focus beam with $n_L\gg 1$ cycles moves at the speed of light and backwards with respect to the beam propagation direction, the effects of the TFL undergo a large enhancement proportional to $n_L$ and may substantially alter the differential emission probability for feasible flying focus pulses.
\end{abstract}

\pacs{12.20.Ds, 41.60.-m}
\maketitle

\section{Introduction}
The emission of radiation by accelerated electric charges is one of the most fundamental processes in physics with applications spanning from high-energy physics and accelerator physics to astrophysics. If charges are accelerated by sufficiently intense electromagnetic fields the emission process can be described theoretically within the framework of strong-field QED, where the influence of the intense background field onto the emission of radiation can be taken into account exactly \cite{Landau_b_4_1982,Fradkin_b_1991,Dittrich_b_1985}. This is achieved by describing the intense electromagnetic field as a given classical background field and by quantizing the electron-positron field in the presence of the background field (Furry picture) \cite{Furry_1951}. By considering for definiteness the emission of radiation by electrons (charge $e<0$ and mass $m$, respectively), the applicability of the Furry picture requires the ability of solving the Dirac equation in the background field analytically, which can be achieved only for particularly symmetric electromagnetic fields like, for example, a plane wave or a Coulomb field \cite{Landau_b_4_1982} (see also the monograph \cite{Bagrov_b_2014}).

Within the Furry picture the transition amplitude $S_{fi}$ of the process of radiation of a single photon by an electron in an external electromagnetic field is expressed as a spacetime integral $S_{fi}=\int d^4x A_{fi}(x)$ \cite{Landau_b_4_1982}, where the complex function $A_{fi}(x)$ depends on the initial and final states of the electron in the background field as well as on the photon state. Thus, the corresponding probability $P_{fi}=|S_{fi}|^2=\int d^4xd^4x'A_{fi}(x)A^*_{fi}(x')$ can be written in the form $P_{fi}=\int d^4x_+ W_{fi}(x_+)$, where
\begin{equation}
\label{W}
W_{fi}(x_+)=\int d^4x_- A_{fi}\left(x_++\frac{x_-}{2}\right) A^*_{fi}\left(x_+-\frac{x_-}{2}\right),
\end{equation}
with $x_{\pm}=(x\pm x')/2^{(1\pm 1)/2}$. The presence of the electron and photon wave functions typically renders the amplitudes $A_{fi}(x)$ oscillating functions. In this respect, the quantities $A_{fi}\left(x_++x_-/2\right) A^*_{fi}\left(x_+-x_-/2\right)$ represent the ``elementary'' contributions to radiation, which typically interfere constructively only within a limited spacetime region of the relative variables $x^{\mu}_-$ for a given spacetime point $x_+$, whereas the contributions from the remaining spacetime volume approximately cancel each other. The region in $x_-$ of constructing interference is known as ``formation region'' and the exact behavior of the integrand in the outer spacetime volume is practically irrelevant for the radiation at $x_+$ [see Eq. (\ref{W})]. For this reason, although being a mathematical construct, the notion of formation region is physically extremely useful in radiation theory, as it allows to assign characteristic values to the coordinates inside the integrals, to simplify the integrands and to gain insights on how the structure of the background field shapes the features of radiation \cite{Ter-Mikaelian_b_1972,Baier_2005}.

If the background field features some spacetime symmetries, the electron states with definite asymptotic momenta have a plane-wave like dependence on the corresponding coordinates. For example, the electron states in a static Coulomb field depend on time as $\exp(-i\varepsilon t/\hbar)$, where $\varepsilon$ is the electron energy. In such cases, the formation volumes may formally extend to infinity along the symmetry directions giving rise to delta functions, which in turn enforce corresponding energy-momentum conservation laws, as energy conservation in a Coulomb field. In this case, one considers the remaining lower-dimensional integral in Eq. (\ref{W}) and introduces corresponding lower-dimensional formation regions. Another example is represented by a plane-wave background field, which depends on a single spacetime variable $\phi=(nx)=ct-\bm{n}\cdot\bm{x}$, where $n^{\mu}=(1,\bm{n})$, with $\bm{n}$ being the unit vector characterizing the propagation direction of the plane wave [the metric tensor $\eta{^{\mu\nu}}=\text{diag}(+1,-1,-1,-1)$ is used throughout]. The process of single-photon emission in an intense plane wave is known as nonlinear single Compton scattering and the corresponding probability is well-known in the literature \cite{Goldman_1964,Nikishov_1964,Brown_1964,Sarachik_1970,Neville_1971,Baier_b_1998,
Ivanov_2004,Boca_2009,Harvey_2009,Mackenroth_2010,Boca_2011,Mackenroth_2011,Seipt_2011,
Seipt_2011b,Dinu_2012,Krajewska_2012,Dinu_2013,Seipt_2013,Krajewska_2014,Wistisen_2014,
Harvey_2015,Seipt_2016,Seipt_2016b,Angioi_2016,Harvey_2016b,Angioi_2018,Di_Piazza_2018,
Alexandrov_2019,Di_Piazza_2019,Ilderton_2019_b} (see also the reviews \cite{Mitter_1975,
Ritus_1985,Ehlotzky_2009,Reiss_2009,Di_Piazza_2012,Dunne_2014}). In this case, indeed, the integral in Eq. (\ref{W}) gives rise to three energy-momentum conservation laws (the transverse momenta $\bm{p}_{\perp}=\bm{p}-(\bm{p}\cdot\bm{n})\bm{n}$ and the quantity $(np)=\varepsilon/c-\bm{p}\cdot\bm{n}$, with $p^{\mu}=(\varepsilon/c,\bm{p}) $, being a generic four-momentum) and one introduces the concept of formation ``phase'' corresponding to the quantity $\phi$ (see, e.g., \cite{Ritus_1985}).

In Refs. \cite{Di_Piazza_2014,Di_Piazza_2015,Di_Piazza_2016,Di_Piazza_2017_b} we have developed a formalism based on the Wentzel–Kramers–Brillouin (WKB) approximation to investigate strong-field QED processes in the presence of tightly focused laser beams in the ultrarelativistic regime. We have found that at the leading order in the ultrarelativistic limit the quantity $(np)$ is still conserved but that the transverse structure of the field may alter the probability. Correspondingly we have introduced the concept of a transverse formation length (TFL) and we have found that it is typically of the order of the Compton wavelength $\lambda_C=\hbar/mc\approx 3.9\times 10^{-11}\;\text{cm}$ \cite{Di_Piazza_2017_b}. Thus, the laser field has been realistically assumed to be constant over the TFL and all interference effects over the TFL have been neglected. As a result, the final emission probability was obtained as the average over the transverse coordinates of the probability in a plane wave, but with the field being locally also dependent on those coordinates \cite{Di_Piazza_2017_b}.

In this paper we show that the interference effects over the TFL alter the emission probability of a single photon in nonlinear Compton scattering by an electron colliding with a so-called  ``flying focus'' (FF) laser beam \cite{Sainte-Marie_2017,Froula_2018,Palastro_2018,Howard_2019,Palastro_2020} (indicated as ``sliding focus'' laser beam in Ref. \cite{Sainte-Marie_2017}). FF laser beams have been realized experimentally and have the unique feature that their focal spot can move virtually at any speed either parallel or anti-parallel with respect to the pulse group velocity \cite{Froula_2018,Howard_2019}. Thus, an ultrarelativistic electron initially counterpropagating with respect to a sufficiently long FF beam with the focus moving at the speed of light also in the opposite direction of the laser propagation direction, would not stay inside the focus for a time corresponding to about two Rayleigh lengths, like in a beam with a fixed focus, but potentially for the whole time duration of the pulse. Now, as we will show analytically below, the TFL is indeed typically of the order of the Compton wavelength but it is also proportional to the square root of the time that the electron spends in the strong field (see below for a technically more precise statement). In this way, the effects of the TFL in the presence of an appropriately prepared long FF beam accumulate and undergo an enhancement by orders of magnitude as compared to a beam with  fixed focus, rendering the observation of the related interference effects in principle feasible. Note that the thoroughly studied failure of the so-called locally-constant field approximation \cite{Baier_1989,Khokonov_2002,Di_Piazza_2007,Wistisen_2015,
Harvey_2015,Dinu_2016,Di_Piazza_2018,Alexandrov_2019,Di_Piazza_2019,Ilderton_2019_b,
Podszus_2019,Ilderton_2019,Raicher_2020} is based on the smallness of the longitudinal formation length (LFL), i.e., the formation time times the speed of light, as compared to the typical wavelength of the laser field. The LFL has a classical counterpart (see, e.g., \cite{Jackson_b_1975}) whereas the TFL is a pure quantum concept, as it is also evinced by its typical value being of the order of $\lambda_C$ \cite{Di_Piazza_2017_b}.

\section{Nonlinear Compton scattering in a focused laser beam}

We consider an arbitrarily focused optical laser beam, described by the four-vector potential $A^{\mu}(x)$, which propagates along the negative $z$ direction, and which is characterized by a central angular frequency $\omega_0$ (corresponding central wavelength $\lambda_0=2\pi/\omega_0$), by an electric field amplitude $E_0$, by a transverse spot radius $\sigma$ (Rayleigh length $l_R=\pi\sigma^2/\lambda_0$), and by a pulse duration $\tau$ (from now on units with $4\pi\epsilon_0=\hbar=c=1$ are employed). In general, it is assumed that $A^{\mu}(x)$ satisfies the free Maxwell's equations with the convenient asymptotic conditions $\lim_{T\to\pm\infty}A^{\mu}(x)=0$, where $T=(t+z)/2$. As it will be clear below, it is convenient to employ light-cone coordinates $T=(t+z)/2$, $\phi=t-z$, and $\bm{x}_{\perp}=(x,y)$ and corresponding light-cone components $v_+=(v_0+v_z)/2$, $v_-=v_0-v_z$, and $\bm{v}_{\perp}=(v_x,v_y)$ for an generic four-vector $v^{\mu}=(v^0,\bm{v})$. By working within the Lorenz gauge $\partial_{\mu}A^{\mu}(x)=0$, the free Maxwell's equations reduce to the free wave equations $\partial_{\mu}\partial^{\mu}A^{\nu}(x)=0$. Concerning in particular the FF beams,  their analytical expression is rather cumbersome \cite{Palastro_2018} and not needed here (see the Appendix A for an exact solution of Maxwell's equations, which can be employed to describe the main features of a FF beam with the focus moving at the speed of light backwards as compared to the laser propagation direction). The $z$-component $v_f$ of the velocity of the focus can be expressed in terms of the focal length $f$ for the central frequency and of the laser chirp parameter $\zeta$ as $v_f=2f/(2f+\zeta\tau^2\omega_0)$ \cite{Palastro_2018}, which indicates as a negative chirp can be chosen to set $v_f=-1$. Optical FF laser beams with intensities $I_0$ close to the relativistic regime ($I_0\sim 10^{18}\;\text{W/cm$^2$}$) are feasible \cite{Palastro_2020} and we assume that the background field is characterized by values of the classical nonlinearity parameter $\xi_0=|e|E_0/m\omega_0=0.75\sqrt{I_0[10^{18}\;\text{W/cm$^2$}]}/\omega_0[\text{eV}]$ of the order of unity.

Passing now to the properties of the incoming electron, we consider an electron with initial four-momentum $p^{\mu}=(\varepsilon,\bm{p})$, with $\varepsilon=\sqrt{m^2+\bm{p}^2}$. As in Refs. \cite{Di_Piazza_2014,Di_Piazza_2015,Di_Piazza_2016,Di_Piazza_2017_b}, the initial electron energy $\varepsilon$ is considered to be the largest dynamical energy in the problem, i.e., $\eta_0=\max{(m,m\xi_0)}/\varepsilon\ll 1$. In addition, the electron is almost counterpropagating with respect to the laser field, i.e., $p_z>0$, $|\bm{p}_{\perp}|\lesssim \max{(m,m\xi_0)}$, and then $p_z\approx \varepsilon$ (we will see below that the condition on $\bm{p}_{\perp}$ is actually more restrictive). Finally, the quantum nonlinearity parameter $\chi_0\approx (2\varepsilon/m)(E_0/E_{cr})\approx 0.057\varepsilon[\text{GeV}]\sqrt{I_0[10^{20}\;\text{W/cm$^2$}]}$ is assumed to be less than or of the order of unity, with $E_{cr}=m^2/|e|\approx 1.3\times 10^{16}\;\text{V/cm}$ being the critical field of QED \cite{Ritus_1985,Baier_b_1998,Di_Piazza_2012}.

Our starting point is the differential emission probability $dP/d\omega$ per unit of emitted photon energy $\omega$ in Eq. (35) in Ref. \cite{Di_Piazza_2017_b}. For the sake of completeness, we provide an easier and more general derivation here.

\subsection{Derivation of the differential emission probability}

First, we recall that in Ref. \cite{Di_Piazza_2017_b} the electron states have been employed, obtained via the WKB method up to the next-to-leading order in Refs. \cite{Di_Piazza_2014,Di_Piazza_2015}. Now, unlike in Ref. \cite{Di_Piazza_2017_b}, we exploit the additional gauge freedom in the Lorenz gauge, to set $A_-(x)=0$ (axial gauge). This choice greatly simplifies already the expressions of the electron in- and out-states because within the WKB method up to the next-to-leading order, the states are independent of $A_+(x)$. For the sake of completeness we report the resulting expression of the electron states:
\begin{align}
\label{in-out_pm_i}
\psi^{(\text{in})}_{p,\sigma}(x)&=e^{iS^{(\text{in})}_p(x)}\bigg[1-\frac{e}{2p_+}\gamma_+\bm{\gamma}_{\perp}\cdot\bm{A}_{\perp}(x)\bigg] \frac{u_{p,\sigma}}{\sqrt{2\varepsilon}},\\
\psi^{(\text{out})}_{p,\sigma}(x)&=e^{iS^{(\text{out})}_p(x)}\bigg[1-\frac{e}{2p_+}\gamma_+\bm{\gamma}_{\perp}\cdot\bm{A}_{\perp}(x)\bigg] \frac{u_{p,\sigma}}{\sqrt{2\varepsilon}},\\
\psi^{(\text{in})}_{-p,-\sigma}(x)&=e^{iS^{(\text{in})}_{-p}(x)}\bigg[1+\frac{e}{2p_+}\gamma_+\bm{\gamma}_{\perp}\cdot\bm{A}_{\perp}(x)\bigg] \frac{u_{-p,-\sigma}}{\sqrt{2\varepsilon}},\\
\label{in-out_pm_f}
\psi^{(\text{out})}_{-p,-\sigma}(x)&=e^{iS^{(\text{out})}_{-p}(x)}\bigg[1+\frac{e}{2p_+}\gamma_+\bm{\gamma}_{\perp}\cdot\bm{A}_{\perp}(x)\bigg] \frac{u_{-p,-\sigma}}{\sqrt{2\varepsilon}},
\end{align}
where $p^{\mu}$ and $\sigma$ correspond to the asymptotic on-shell four-momentum and spin quantum number, respectively, where
\begin{align}
\label{S_in}
S^{(\text{in})}_{\pm p}(x)&=\mp(p_+\phi+p_-T-\bm{p}_{\perp}\cdot\bm{x}_{\perp})+\frac{1}{p_+}\int_{-\infty}^Td\bar{T} \left[e\bm{p}_{\perp}\cdot\bm{A}_{\perp}(\bar{x})\mp
\frac{1}{2}e^2\bm{A}^2_{\perp}(\bar{x})\right],\\
\label{S_out}
S^{(\text{out})}_{\pm p}(X)&=\mp(p_+\phi+p_-T-\bm{p}_{\perp}\cdot\bm{x}_{\perp})-\frac{1}{p_+}\int_T^{\infty}d\bar{T} \left[e\bm{p}_{\perp}\cdot\bm{A}_{\perp}(\bar{x})\mp
\frac{1}{2}e^2\bm{A}^2_{\perp}(\bar{x})\right],
\end{align}
with $\bar{x}=(\bar{T},\bm{x}_{\perp},\phi)$, and where $u_{\pm p,\pm \sigma}$ are the constant bi-spinors with positive and negative energy \cite{Landau_b_4_1982}. The achieved simplification can be appreciated by comparing the above equations with Eqs. (22)-(29) in Ref. \cite{Di_Piazza_2015}, by noticing that in the present gauge and with the asymptotic conditions on the four-vector potential, we simply have $\bm{\mathcal{A}}^{(\text{in})}_{\perp}(x)=\bm{\mathcal{A}}^{(\text{out})}_{\perp}(x)=\bm{A}_{\perp}(x)=-\int_{-\infty}^Td\bar{T}[\bm{E}_{\perp}(\bar{x})+\bm{z}\times\bm{B}_{\perp}(\bar{x})]$, with $(\bm{E}(x),\bm{B}(x))$ being the background electromagnetic field (note that, unlike in Ref. \cite{Di_Piazza_2015}, here we use units with $4\pi\epsilon_0=\hbar=c=1$ and we explicitly indicate the dependence of the field on the light-cone variable $\phi$).

We pass now to the description of nonlinear single Compton scattering. It is convenient initially to assume that the incoming electron is described by the wave packet
\begin{equation}
\Psi^{(\text{in})}_{p,\sigma}(x)=\int\frac{d^3\bm{q}}{(2\pi)^3}\rho_p(\bm{q})\psi^{(\text{in})}_{p,\sigma}(x),
\end{equation}
with fixed spin quantum number $\sigma$ and momentum distribution $\rho_p(\bm{q})$ well peaked around the momentum $\bm{p}$, corresponding to the energy $\varepsilon=\sqrt{m^2+\bm{p}^2}$. Also, the central momentum and the distribution $\rho_p(\bm{q})$ are assumed to correspond to an ultrarelativistic electron almost counterpropagating with respect to the laser field, according to the general method developed in Refs. \cite{Di_Piazza_2014,Di_Piazza_2015}. Finally, the wave packet is assumed to be normalized to unity as
\begin{equation}
\int \frac{d^3q}{(2\pi)^3}|\rho_p(\bm{q})|^2=1.
\end{equation}
Here, we observe that the states $\psi^{(\text{in})}_{p,\sigma}(x)$ are approximated solutions of the Dirac equation valid for ultrarelativistic energies and up to terms scaling as $1/p_+\approx 1/\varepsilon$. By using the Dirac equation for a generic state $\psi^{(\text{in})}_{p,\sigma}(x)$ and for its Hermitian conjugated and by imposing periodic boundary conditions on a finite volume $V$, it is easy to show that
\begin{equation}
\frac{d}{dt}\int_Vd^3\bm{x}\,\psi^{(\text{in})\dag}_{p',\sigma'}(x)\psi^{(\text{in})}_{p,\sigma}(x)=0+O(1/\varepsilon^2),
\end{equation}
and that the orthogonality and normalization properties of the $\psi^{(\text{in})}_{p,\sigma}(x)$ are the same as for the free states apart from terms scaling at least as $1/\varepsilon^2$. In the limit $V\to\infty$ we obtain 
\begin{equation}
\int d^3\bm{x}\,\psi^{(\text{in})\dag}_{p',\sigma'}(x)\psi^{(\text{in})}_{p,\sigma}(x)=(2\pi)^3\delta(\bm{p}-\bm{p'})\delta_{\sigma,\sigma'}+O(1/\varepsilon^2),
\end{equation}
and analogous for the relations involving the negative-energy states.

Now, we assume that the final electron has on-shell four-momentum $p^{\prime\mu}=(\varepsilon',\bm{p}')$ and spin quantum number $\sigma'$. Analogously the emitted photon has on-shell four-momentum $k^{\mu}=(\omega,\bm{k})$ and (linear) polarization $l$ (polarization four-vector $e^{\mu}_{k,l}$). The leading-order $S$-matrix element of nonlinear single Compton scattering in the Furry picture reads \cite{Furry_1951,Landau_b_4_1982}
\begin{equation}
\label{S_C}
S_{fi}=-ie\sqrt{4\pi}\int d^4x\,\bar{\psi}^{(\text{out})}_{p',\sigma'}(x)\frac{\hat{e}_{k,l}}{\sqrt{2\omega}}e^{i(kx)}\Psi^{(\text{in})}_{p,\sigma}(x).
\end{equation}
Since the momentum distribution function $\rho_p(\bm{q})$ is well peaked around the momentum $\bm{p}$, which corresponds to the on-shell four-momentum $p^{\mu}=(\varepsilon,\bm{p})$, we can approximate the $S$-matrix element in Eq. (\ref{S_C}) as
\begin{equation}
S_{fi}\approx \int d^4x\,\tilde{\rho}_p(x)M_{fi,p}(x).
\end{equation}
where
\begin{equation}
M_{fi,p}(x)=-ie\sqrt{4\pi}\bar{\psi}^{(\text{out})}_{p',\sigma'}(x)\frac{\hat{e}_{k,l}}{\sqrt{2\omega}}e^{i(kx)}\psi^{(\text{in})}_{p,\sigma}(x).
\end{equation}
is the matrix element corresponding to an electron with four-momentum $p^{\mu}$ and where
\begin{equation}
\label{rho}
\tilde{\rho}_p(x)=\int\frac{d^3\bm{q}}{(2\pi)^3}\rho_p(\bm{q}) e^{i[S_q^{(\text{in})}(x)-S_p^{(\text{in})}(x)]},
\end{equation}
is the spin-independent amplitude of the wave packet in configuration space. The expression in Eq. (\ref{rho}) reminds that care has to be taken to treat the oscillating exponential functions also if the function $\rho_p(\bm{q})$ is well peaked around the momentum $\bm{p}$ (see also Ref.  \cite{Bragin_2020}).

The corresponding differential probability of the process by averaging (summing) over the initial (final) discrete quantum numbers is given by 
\begin{equation}
dP=\frac{d^3\bm{k}}{(2\pi)^3} \frac{d^3\bm{p}'}{(2\pi)^3} \frac{1}{2}\sum_{l,\sigma,\sigma'}|S_{fi}|^2,
\end{equation}
and, by following exactly the same steps as in Ref. \cite{Di_Piazza_2017_b}, we arrive to the expression
\begin{equation}
\label{dN_1_C}
\begin{split}
dP=&\frac{\pi\alpha}{\omega\varepsilon\varepsilon'}\frac{d\varepsilon'}{2\pi}\frac{d^2\bm{p}'_{\perp}}{(2\pi)^2}\frac{d\omega}{2\pi}\frac{d^2\bm{k}_{\perp}}{(2\pi)^2}\int d^4xd^4x'\,\tilde{\rho}_p(x)\tilde{\rho}^*_p(x')e^{i[\Phi_C(x)-\Phi_C(x')]}\left\{m^2\left(\frac{\varepsilon'}{\varepsilon}+\frac{\varepsilon}{\varepsilon'}-4\right)\right.\\
&+\frac{\varepsilon'}{\varepsilon}\bm{p}^2_{\perp}-2\bm{p}_{\perp}\cdot\bm{p}'_{\perp}+\frac{\varepsilon}{\varepsilon'}\bm{p}^{\prime\, 2}_{\perp}+e\frac{\omega}{\varepsilon\varepsilon'}(\varepsilon'\bm{p}_{\perp}-\varepsilon\bm{p}'_{\perp})\cdot[\bm{A}_{\perp}(x)+\bm{A}_{\perp}(x')]\\
&\left.-e^2\left[\bm{A}^2_{\perp}(x)+\bm{A}^2_{\perp}(x')-\left(\frac{\varepsilon'}{\varepsilon}+\frac{\varepsilon}{\varepsilon'}\right)\bm{A}_{\perp}(x)\cdot\bm{A}_{\perp}(x')\right]\right\},
\end{split}
\end{equation}
where $\alpha=e^2\approx 1/137$ is the fine-structure constant. Here, we have used the fact that within the first-order WKB approach we can approximate $p_+\approx\varepsilon$, $p'_+\approx\varepsilon'$ and $k_+\approx\omega$, and we have introduced the phase 
\begin{equation}
\begin{split}
\Phi_C(x)=&(\varepsilon'+\omega-\varepsilon)\phi+\left(\frac{m^2+\bm{p}^{\prime 2}_{\perp}}{2\varepsilon'}+\frac{\bm{k}_{\perp}^2}{2\omega}-\frac{m^2+\bm{p}_{\perp}^2}{2\varepsilon}\right)T-(\bm{p}'_{\perp}+\bm{k}_{\perp}-\bm{p}_{\perp})\cdot\bm{x}_{\perp}\\
&+e\frac{\bm{p}'_{\perp}}{\varepsilon'}\cdot\int_T^{\infty}d\bar{T} \bm{A}_{\perp}(\bar{x})+e\frac{\bm{p}_{\perp}}{\varepsilon}\cdot\int_{-\infty}^Td\bar{T} \bm{A}_{\perp}(\bar{x})-\frac{1}{\varepsilon'}\frac{e^2}{2}\int_T^{\infty}d\bar{T} \bm{A}^2_{\perp}(\bar{x})\\
&-\frac{1}{\varepsilon}\frac{e^2}{2}\int_{-\infty}^Td\bar{T} \bm{A}^2_{\perp}(\bar{x}).
\end{split}
\end{equation}
The last two equations exactly correspond to Eqs. (31)-(32) in Ref. \cite{Di_Piazza_2017_b}, but where we have still not taken the integral in the variable $\phi$ (which will enforce the energy conservation $\varepsilon=\varepsilon'+\omega$).

Before continuing with the computation, we observe that for a sufficiently narrow wave packet in momentum space, we can expand the exponent in Eq. (\ref{rho}) up to linear terms:
\begin{equation}
\label{rho_1}
\tilde{\rho}_p(x)=\int\frac{d^3\bm{q}}{(2\pi)^3}\rho_p(\bm{q}) e^{i[\bm{\nabla}_{\bm{p}}S_p^{(\text{in})}(x)]\cdot(\bm{q}-\bm{p})}.
\end{equation}
According to the general theory as presented, e.g., in Ref. \cite{Goldberger_b_1964}, this approximation amounts to neglect the spreading of the wave packet. Also, in the WKB approach followed in Ref. \cite{Di_Piazza_2017_b} and here, the quantity $S_p^{(\text{in})}(x)$ is the action corresponding to the electron trajectory in the external field with the momentum $(p_+,\bm{p}_{\perp})$ at asymptotic early time $T\to-\infty$ (when the electron moves freely outside the field) and with position $(\bm{x}_{\perp},\phi)$ at the generic finite time $T$. By assuming that, at a sufficiently early time $T_0$ such that the integral in the action $S_p^{(\text{in})}(x)$ in Eq. (\ref{S_in}) can be neglected, the position of the electron (outside the field) corresponds to the coordinates $\bm{x}_{0,\perp}$ and $\phi_0$, then the asymptotic free trajectory of the electron can be parametrized as $\bm{x}_{\perp}=\bm{x}_{0,\perp}+(\bm{p}_{\perp}/\varepsilon)(T-T_0)$ and $\phi=\phi_0+(p_-/\varepsilon)(T-T_0)$, with $p_-=(m^2+\bm{p}_{\perp}^2)/2\varepsilon$. Therefore, according to the general theory of mechanical systems \cite{Goldstein_b_2002}, the quantities $\bm{\nabla}_{\bm{p}_{\perp}}S_p^{(\text{in})}(x)$ and $\partial_{p_+}S_p^{(\text{in})}(x)$ correspond to the quantities $\bm{x}_{0,\perp}-(\bm{p}_{\perp}/\varepsilon)T_0$ and $-\phi_0+(p_-/\varepsilon)T_0$, respectively. Thus, as expected, if the function $\tilde{\rho}_p(x)$ is centered at a given early asymptotic time $T_0$ around the point $(\bm{x}_{0,\perp},\phi_0)$, then Eq. (\ref{rho_1}) implies that at a generic late time $T$ it will be centered around the position of the electron at that time on the corresponding classical trajectory in the external field.

By passing in Eq. (\ref{dN_1_C}) to the centered and the relative variables $x_+=(x+x')/2$ and $x_-=x-x'$, respectively, we notice that the relative coordinate $\phi_-$ can be integrated out as the field can be evaluated everywhere at the centered coordinate $\phi_+$. The reason is that, since $\phi_-=t-z$ and the electron propagates with ultrarelativistic velocity along the positive $z$ axis, the formation length in $\phi_-$ scales as the inverse of the square of the electron energy \cite{Di_Piazza_2014,Di_Piazza_2015,Di_Piazza_2016,Di_Piazza_2017_b}. Thus, within the first-order WKB approximation, the dependence of the field on the quantity $\phi_-$ can be neglected and the integral over $\phi_-$ provides the energy conservation condition $\varepsilon=\varepsilon'+\omega$. Moreover, since the function $\tilde{\rho}_p(x)$ does not depend on the transverse momenta of the final electron and of the photon, the corresponding integrals can be taken as in Ref. \cite{Di_Piazza_2017_b} and we obtain
\begin{equation}
\label{dN_domega_App}
\begin{split}
&\frac{dP}{d\omega}=-\frac{\alpha}{8\pi^2\varepsilon}\int d^4x_+\int\frac{dT_-d^2\bm{x}_{-,\perp}}{T^2_-}\,\tilde{\rho}_p(x)\tilde{\rho}^*_p(x')\\
&\qquad\times\exp\left\langle i\frac{T_-}{2}\left\{\frac{m^2\omega}{\varepsilon\varepsilon'}-\frac{\varepsilon}{T^2_-}\left[\bm{x}_{-,\perp}-\frac{T_-}{\varepsilon}(\bm{p}_{\perp}-\bm{I}_{\text{in}})\right]^2+\frac{1}{\varepsilon}(\bm{I}^2_{\text{in}}-J_{\text{in}})-\frac{1}{\varepsilon'}(\bm{I}^2_{\text{out}}-J_{\text{out}})\right\}\right\rangle\\
&\qquad\times\left\langle m^2\left(\frac{\varepsilon'}{\varepsilon}+\frac{\varepsilon}{\varepsilon'}-4\right)+\frac{2i\varepsilon}{T_-}+\frac{\varepsilon'}{\varepsilon}\bigg\{\frac{\varepsilon}{T_-}\bm{x}_{-,\perp}+\frac{\varepsilon}{\varepsilon'}\bm{I}_{\text{out}}-\frac{\omega}{2\varepsilon'}\left[\bm{\mathcal{A}}_{\perp}(x)+\bm{\mathcal{A}}_{\perp}(x')\right]\bigg\}^2\right.\\
&\qquad\left.-\frac{(\varepsilon+\varepsilon')^2}{4\varepsilon\varepsilon'}\left[\bm{\mathcal{A}}_{\perp}(x)-\bm{\mathcal{A}}_{\perp}(x')\right]^2\right\rangle,
\end{split}
\end{equation}
where $x_{\pm}=(T_{\pm},\bm{x}_{\pm,\perp},\phi_+)=(x\pm x')/2^{(1\pm 1)/2}$, with $x=(T,\bm{x}_{\perp},\phi_+)$ and $x'=(T',\bm{x}'_{\perp},\phi_+)$, where
\begin{align}
\label{I_in_out}
\bm{I}_{\text{in}/\text{out}}&=\frac{1}{T_-}\left[\int_{\mp\infty}^Td\tilde{T} \bm{\mathcal{A}}_{\perp}(\tilde{x})-\int_{\mp\infty}^{T'}d\tilde{T}' \bm{\mathcal{A}}_{\perp}(\tilde{x}')\right],\\
\label{J_in_out}
J_{\text{in}/\text{out}}&=\frac{1}{T_-}\left[\int_{\mp\infty}^Td\tilde{T} \bm{\mathcal{A}}^2_{\perp}(\tilde{x})-\int_{\mp\infty}^{T'}d\tilde{T}' \bm{\mathcal{A}}^2_{\perp}(\tilde{x}')\right],
\end{align}
with $\tilde{x}=(\tilde{T},\bm{x}_{\perp},\phi_+)$ and $\tilde{x}'=(\tilde{T}',\bm{x}'_{\perp},\phi_+)$, and where $\bm{\mathcal{A}}_{\perp}(x)=e\bm{A}_{\perp}(x)$.

Finally, if the incoming electron is in the definite momentum state corresponding to the central momentum $\bm{p}$, i.e., for $\rho_p(\bm{q})=(2\pi)^3\delta^3(\bm{q}-\bm{p})\sqrt{\rho_0}$, with $\rho_0$ being a constant electron spatial density, Eq. (\ref{dN_domega_App}) can be written in the form $dP/d\omega=\int d^4x_+\,dW(x_+)/d\omega$, where,
\begin{equation}
\label{dN_domega}
\begin{split}
\frac{dW(x_+)}{d\omega}&=-\frac{\alpha\rho_0}{8\pi^2\varepsilon}\int\frac{dT_-d^2\bm{x}_{-,\perp}}{T^2_-}e^{i\Phi}\left\langle m^2\left(\frac{\varepsilon'}{\varepsilon}+\frac{\varepsilon}{\varepsilon'}-4\right)+\frac{2i\varepsilon}{T_-}\right.\\
&\quad+\frac{\varepsilon'}{\varepsilon}\bigg\{\frac{\varepsilon}{T_-}\bm{x}_{-,\perp}-\bm{p}_{\perp}+\frac{\varepsilon}{\varepsilon'}\bm{I}_{\text{out}}-\frac{\omega}{2\varepsilon'}\left[\bm{\mathcal{A}}_{\perp}(x)+\bm{\mathcal{A}}_{\perp}(x')\right]\bigg\}^2\\
&\quad\left.-\frac{(\varepsilon+\varepsilon')^2}{4\varepsilon\varepsilon'}\left[\bm{\mathcal{A}}_{\perp}(x)-\bm{\mathcal{A}}_{\perp}(x')\right]^2\right\rangle,
\end{split}
\end{equation}
with
\begin{equation}
\label{Phi}
\Phi=\frac{T_-}{2}\left\{\frac{m^2\omega}{\varepsilon\varepsilon'}-\frac{\varepsilon}{T^2_-}\left[\bm{x}_{-,\perp}-\frac{T_-}{\varepsilon}(\bm{p}_{\perp}-\bm{I}_{\text{in}})\right]^2+\frac{1}{\varepsilon}(\bm{I}^2_{\text{in}}-J_{\text{in}})-\frac{1}{\varepsilon'}(\bm{I}^2_{\text{out}}-J_{\text{out}})\right\},
\end{equation}
which exactly corresponds to Eq. (35) in Ref. \cite{Di_Piazza_2017_b} [note an evident misprint in the pre-exponent of Eq. (35) in Ref. \cite{Di_Piazza_2017_b}, where the term $2i\omega/T_-$ should rather read $2i\varepsilon/T_-$, as it is clear from the previous Eq. (33) there].

\subsection{Analysis of the transverse formation length of nonlinear Compton scattering}

In order to investigate the properties of the TFL, we have to analyze the integral over $\bm{x}_{-,\perp}$ in Eq. (\ref{dN_domega}) and we recall that the external field obviously also depends on the quantity $\bm{x}_{-,\perp}$. The strategy is to pass from the variable $\bm{x}_{-,\perp}$ to the variable $\bm{\rho}_{\perp}=\bm{x}_{-,\perp}-\bm{R}_{\perp}$, where $\bm{R}_{\perp}$ is a quantity independent of $\bm{\rho}_{\perp}$ and to be chosen in such a way that the following two conditions are fulfilled: 1) the resulting integral over $\bm{\rho}_{\perp}$ is formed around a region much smaller than the laser spot radius $\sigma$, i.e., the typical transverse length where the field changes significantly; 2) the terms linear in $\bm{\rho}_{\perp}$ in the phase $\Phi$ in Eq. (\ref{Phi}) resulting after expanding the fields in $\Phi$ for small values of $|\bm{\rho}_{\perp}|$ vanish. We will see below that these requirements can be self-consistently  fulfilled and that the vector $\bm{R}_{\perp}$ is related to the trajectory of the electron on the transverse plane. We also anticipate that, as expected, the TFL $l_{\perp}$ will correspond to the region where the integral in $\bm{\rho}_{\perp}$ is formed. It is sufficient here to carry out the expansion of the external field up to the first order in $\bm{\rho}_{\perp}$ in Eq. (\ref{Phi}), which will appear within the operator $\delta_{\perp}=\bm{\rho}_{\perp}\cdot\bm{\nabla}_{\perp}$, with $\bm{\nabla}_{\perp}=\partial/\partial \bm{x}_{+,\perp}$. By indicating as $\Phi^{(1)}$ the corresponding phase up to the first order in $\delta_{\perp}$, it is easily shown that
\begin{equation}
\label{Phi_1}
\begin{split}
\Phi^{(1)}&=\frac{T_-}{2}\bigg\{\frac{m^2\omega}{\varepsilon\varepsilon'}-\frac{\varepsilon}{T^2_-}\left[\bm{R}_{\perp}-\frac{T_-}{\varepsilon}(\bm{p}_{\perp}-\bm{I}^{(-)}_{\text{in}})\right]^2+\frac{1}{\varepsilon}(\bm{I}^{(-)\,2}_{\text{in}}-J^{(-)}_{\text{in}})-\frac{1}{\varepsilon'}(\bm{I}^{(-)\,2}_{\text{out}}-J^{(-)}_{\text{out}})\\
&\quad-\frac{\varepsilon}{T^2_-}\bm{\rho}_{\perp}^2-\frac{1}{T_-}\bm{\rho}_{\perp}\cdot\delta_{\perp}\bm{I}^{(+)}_{\text{in}}-\frac{1}{T_-}\left(\frac{2\varepsilon}{T_-}\bm{\rho}_{\perp}+\delta_{\perp}\bm{I}^{(+)}_{\text{in}}\right)\cdot\left[\bm{R}_{\perp}-\frac{T_-}{\varepsilon}(\bm{p}_{\perp}-\bm{I}^{(-)}_{\text{in}})\right]\\
&\quad+\frac{1}{\varepsilon}\bigg(\bm{I}^{(-)}_{\text{in}}\cdot\delta_{\perp}\bm{I}^{(+)}_{\text{in}}-\frac{\delta_{\perp}}{2}J^{(+)}_{\text{in}}\bigg)-\frac{1}{\varepsilon'}\bigg(\bm{I}^{(-)}_{\text{out}}\cdot\delta_{\perp}\bm{I}^{(+)}_{\text{out}}-\frac{\delta_{\perp}}{2}J^{(+)}_{\text{out}}\bigg)\bigg\}.
\end{split}
\end{equation}
Here, we have introduced the quantities
\begin{align}
\bm{I}^{(\text{sgn}(s))}_{\text{in}/\text{out}}&=\frac{1}{T_-}\left[\int_{\mp\infty}^Td\tilde{T} \bm{\mathcal{A}}_{\perp}(\tilde{X})+s\int_{\mp\infty}^{T'}d\tilde{T}' \bm{\mathcal{A}}_{\perp}(\tilde{X}')\right],\\
J^{(\text{sgn}(s))}_{\text{in}/\text{out}}&=\frac{1}{T_-}\left[\int_{\mp\infty}^Td\tilde{T} \bm{\mathcal{A}}^2_{\perp}(\tilde{X})+s\int_{\mp\infty}^{T'}d\tilde{T}' \bm{\mathcal{A}}^2_{\perp}(\tilde{X}')\right],
\end{align}
where $\tilde{X}=(\tilde{T},\bm{x}_{+,\perp}+\bm{R}_{\perp}/2,\phi_+)$ and $\tilde{X}'=(\tilde{T}',\bm{x}_{+,\perp}-\bm{R}_{\perp}/2,\phi_+)$, and $s=\pm 1$. According to the above discussion, the vector $\bm{R}_{\perp}$ is determined by imposing that the linear terms in $\bm{\rho}_{\perp}$ in $\Phi^{(1)}$ identically vanish. This condition, together with an inspection at the terms in Eq. (\ref{Phi_1}) quadratic in $\bm{\rho}_{\perp}$, implies that the integral in $\bm{\rho}_{\perp}$ is formed within the region $|\bm{\rho}_{\perp}|\lesssim \sqrt{2|T_-|/\varepsilon}$ [note that the second term quadratic in $\bm{\rho}_{\perp}$ is in order of magnitude $|\bm{\rho}_{\perp}\cdot\delta_{\perp}\bm{I}^{(+)}_{\text{in}}|\lesssim (\lambda_0/\sigma)(m\xi_0/\varepsilon)\ll 1$] \footnote{A technical note is in order here. The fact that the TFL is determined by the condition $|\bm{\rho}_{\perp}|\lesssim \sqrt{2|T_-|/\varepsilon}$ seems to contradict the discussion in the introduction because this condition would remain unchanged even if the external field would not depend on the transverse coordinates, as in the plane-wave case. The reason for this apparent contradiction is that, for the sake of convenience in the general case of a field also dependent on the transverse coordinates, the integral over the final momentum $\bm{p}'_{\perp}$ has been taken already in Eq. (\ref{dN_domega}) before the integral over $\bm{x}_{-,\perp}$. However, if the field is independent of $\bm{x}_{\perp}$, one can directly take the integral over these variables at the amplitude level and obtain the conservation delta functions on the transverse momenta, corresponding to an infinite formation region over the transverse plane.}. Thus, we find $l_{\perp}=\sqrt{2|T_-|/\varepsilon}=2\lambda_C\sqrt{\omega_0|T_-|\xi_0/\chi_0}$. Now, since $T=(t+z)/2$, we can identify $|T_-|$ with the formation time or, in our units, with the LFL. Now, if $\xi_0\gg 1$ the LFL in nonlinear Compton scattering is typically much smaller than the laser wavelength \cite{Ritus_1985}. On the contrary, if $\xi_0\lesssim 1$ the radiation is formed over the whole pulse length. Here, one has to point out that in the case of a monochromatic or quasi monochromatic pulse, the motion of the electron is periodic or quasi periodic, such that the amplitude of the process can be written (approximately in the quasi monochromatic case) as the integral over a laser wavelength times the number of wavelengths and the LFL is identified with the wavelength itself. However, the important point here is that the integral over $T_-$ receives contributions from the overall time that the electron spends inside the field. In this respect, in order to estimate the size of the effects of the TFL on the differential emission probability, we assume from now on that $\xi_0\lesssim 1$ and we can estimate $|T_-|\sim \min(2l_R,\tau)$ for a pulse with fixed focus and $|T_-|\sim \tau$ for a FF pulse with the focus moving at the speed of light in the same direction of the electron. Thus, for a pulse with fixed focus $l_{\perp}\lesssim 4\pi\lambda_C(\sigma/\lambda_0)\sqrt{\xi_0/\chi_0}$ and then $l_{\perp}\ll \sigma$ for any realistic optical laser. On the contrary, in the case of a FF pulse, it is $l_{\perp}\sim\lambda_C\sqrt{8\pi n_L\xi_0/\chi_0}$, where $n_L=\tau/\lambda_0$ is the number of cycles in the pulse, which shows the possible large enhancement of the effects of the TFL in this case for $n_L\gg 1$ (in some sense a FF beam behaves as a conventional focused beam but with a potentially extremely long focal region). Thus, we can expect that in the case of a FF pulse, TFL effects can be enhanced as compared with a traditional beam. Nevertheless, as it is required by our perturbative approach, we continue assuming that also in the case of a FF pulse it is $l_{\perp}\ll \sigma$. At this point, we have to determine the vector $\bm{R}_{\perp}$, which, according to the above discussion and up to first order in the transverse field derivatives, has to fulfill the nonlinear equation [see the second line of Eq. (\ref{Phi_1})]
\begin{equation}
\label{R}
\begin{split}
\bm{R}_{\perp}&=\frac{T_-}{\varepsilon}\bigg[\bm{p}_{\perp}-\bm{I}^{(-)}_{\text{in}}+\frac{T_-}{\varepsilon}\bigg(I^{(-)}_{\text{in},j}\bm{\nabla}_{\perp}I^{(+)}_{\text{in},j}-\frac{1}{2}\bm{\nabla}_{\perp}J^{(+)}_{\text{in}}\bigg)\\
&\quad-\frac{T_-}{\varepsilon'}\bigg(I^{(-)}_{\text{out},j}\bm{\nabla}_{\perp}I^{(+)}_{\text{out},j}-\frac{1}{2}\bm{\nabla}_{\perp}J^{(+)}_{\text{out}}\bigg)\bigg],
\end{split}
\end{equation}
where a sum over $j=x,y$ is understood (recall that the field also depends on $\bm{R}_{\perp}$). The physical interpretation of this equation will allow us to further simplify it. In fact, the vector $\bm{R}_{\perp}$ describes the transverse trajectory of the electron inside the field, with the first term in Eq. (\ref{R}) corresponding to the free component of the motion if $\bm{p}_{\perp}\neq \bm{0}$, the second term corresponding to the oscillatory motion due to the laser field, and the remaining terms corresponding to the corrections to this motion due to the non-trivial transverse structure of the field. Now, we notice that $(|T_-|/\varepsilon)|\bm{I}^{(-)}_{\text{in}}|\sim\lambda_0m\xi_0/\varepsilon\ll \lambda_0$ and the corrections induced by this term can be ignored for a leading-order result in $\eta_0$. Correspondingly, one can also approximate $\bm{R}_{\perp}\approx \bm{r}_{\perp}=(\bm{p}_{\perp}/\varepsilon)T_-$ inside the fields in Eq. (\ref{R}), which becomes an explicit expression for $\bm{R}_{\perp}$ as the fields are evaluated at $\tilde{X}=(\tilde{T},\bm{x}_{+,\perp}+\bm{r}_{\perp}/2,\phi_+)$ and $\tilde{X}'=(\tilde{T}',\bm{x}_{+,\perp}-\bm{r}_{\perp}/2,\phi_+)$. A consistent computation of the first-order correction of the differential probability, however, requires to keep all the terms in $\bm{R}_{\perp}$ in Eq. (\ref{R}) because $\bm{x}_{-,\perp}$ also appears explicitly in the pre-exponent and not only inside the laser field [see Eq. (\ref{dN_domega})].

At this point, it is straightforward to compute the leading-order expression $dW_0(x_+)/d\omega$ and the first-order correction $dW_1(x_+)/d\omega$:
\begin{align}
\label{dP_domega_0}
\begin{split}
\frac{dW_0(x_+)}{d\omega}&=\frac{i\alpha\rho_0}{4\pi\varepsilon^2}\int\frac{dT_-}{T_-}e^{i\Phi_0}\bigg[m^2\left(\frac{\varepsilon'}{\varepsilon}+\frac{\varepsilon}{\varepsilon'}-4\right)+\frac{2i\omega}{T_-}+\frac{\varepsilon'}{\varepsilon}\bigg(\bm{I}^{(-)}_{\text{in}}-\frac{\varepsilon}{\varepsilon'}\bm{I}^{(-)}_{\text{out}}+\frac{\omega}{2\varepsilon'}\bm{\mathcal{A}}_{\perp}^{(+)}\bigg)^2\\
&\quad-\frac{(\varepsilon+\varepsilon')^2}{4\varepsilon\varepsilon'}\bm{\mathcal{A}}_{\perp}^{(-)\,2}\bigg],
\end{split}\\
\label{dP_domega_1}
\begin{split}
\frac{dW_1(x_+)}{d\omega}&=\frac{i\alpha\rho_0}{4\pi\varepsilon^2}\int dT_-\,e^{i\Phi_0}\bigg\langle \frac{\varepsilon'}{\varepsilon}\bigg\{\bigg(\bm{I}^{(+)}_{\text{in}}-\frac{\varepsilon}{\varepsilon'}\bm{I}^{(+)}_{\text{out}}+\frac{\omega}{2\varepsilon'}\bm{\mathcal{A}}^{(-)}_{\perp}\bigg)\\
&\quad\cdot\bigg[\frac{1}{\varepsilon}\bigg(I^{(-)}_{\text{in},j}\bm{\nabla}_{\perp}I^{(-)}_{\text{in},j}-\frac{\bm{\nabla}_{\perp}}{2}J^{(-)}_{\text{in}}\bigg)-\frac{1}{\varepsilon'}\bigg(I^{(-)}_{\text{out},j}\bm{\nabla}_{\perp}I^{(-)}_{\text{out},j}-\frac{\bm{\nabla}_{\perp}}{2}J^{(-)}_{\text{out}}\bigg)\bigg]\\
&\quad-\bigg(\bm{I}^{(-)}_{\text{in}}-\frac{\varepsilon}{\varepsilon'}\bm{I}^{(-)}_{\text{out}}+\frac{\omega}{2\varepsilon'}\bm{\mathcal{A}}^{(+)}_{\perp}\bigg)\cdot\bigg[\frac{1}{\varepsilon}\bigg(I^{(-)}_{\text{in},j}\bm{\nabla}_{\perp}I^{(+)}_{\text{in},j}-\frac{\bm{\nabla}_{\perp}}{2}J^{(+)}_{\text{in}}\bigg)\\
&\quad-\frac{1}{\varepsilon'}\bigg(I^{(-)}_{\text{out},j}\bm{\nabla}_{\perp}I^{(+)}_{\text{out},j}-\frac{\bm{\nabla}_{\perp}}{2}J^{(+)}_{\text{out}}\bigg)\bigg]\bigg\}-\frac{\omega}{\varepsilon}\bm{I}^{(+)}_{\text{in}}\cdot\bigg[\frac{1}{\varepsilon}\bigg(I^{(-)}_{\text{in},j}\bm{\nabla}_{\perp}I^{(-)}_{\text{in},j}-\frac{\bm{\nabla}_{\perp}}{2}J^{(-)}_{\text{in}}\bigg)\\
&\quad-\frac{1}{\varepsilon'}\bigg(I^{(-)}_{\text{out},j}\bm{\nabla}_{\perp}I^{(-)}_{\text{out},j}-\frac{\bm{\nabla}_{\perp}}{2}J^{(-)}_{\text{out}}\bigg)\bigg]\bigg\rangle,
\end{split}
\end{align}
where 
\begin{equation}
\label{Phi_0}
\Phi_0=\frac{T_-}{2}\left[\frac{m^2\omega}{\varepsilon\varepsilon'}+\frac{1}{\varepsilon}(\bm{I}^{(-)\,2}_{\text{in}}-J^{(-)}_{\text{in}})-\frac{1}{\varepsilon'}(\bm{I}^{(-)\,2}_{\text{out}}-J^{(-)}_{\text{out}})\right]
\end{equation}
and where $\bm{\mathcal{A}}^{(\pm)}_{\perp}=\bm{\mathcal{A}}_{\perp}(T,\bm{x}_{+,\perp}+\bm{r}_{\perp}/2,\phi_+)\pm\bm{\mathcal{A}}_{\perp}(T',\bm{x}_{+,\perp}-\bm{r}_{\perp}/2,\phi_+)$. First, we notice that the expression of $dP_0/d\omega=\int d^4x_+\,dW_0(x_+)/d\omega$ reduces to the corresponding quantity in Ref. \cite{Di_Piazza_2017_b} in the case $\bm{p}_{\perp}=\bm{0}$ considered there. Also, as we have mentioned, we are interested in the case of long pulses such that the quantities $\bm{I}^{(\pm)}_{\text{in/out}}$ are typically much smaller than $\bm{\mathcal{A}}^{(\pm)}_{\perp}$. Instead, the quantities $J^{(\pm)}_{\text{in/out}}$ contain integrals of the square of the fields, which accumulate, and then we conclude that for long pulses
\begin{equation}
\label{dW_domega_1}
\frac{dW_1(x_+)}{d\omega}\approx\frac{i\alpha\rho_0\omega}{16\pi\varepsilon^3}\int dT_-\,e^{i\Phi_0}\sum_{s=-1,+1}s\bm{\mathcal{A}}^{(\text{sgn}(s))}_{\perp}\cdot\bm{\nabla}_{\perp}J^{(\text{sgn}(s))},
\end{equation}
where $J^{(\pm)}=J^{(\pm)}_{\text{in}}/\varepsilon-J^{(\pm)}_{\text{out}}/\varepsilon'$. This equation shows that in the case of a FF with pulse duration $\tau$ and at $\omega\sim \varepsilon\sim\varepsilon'$, the correction is about
\begin{equation}
\label{theta}
\theta=\frac{1}{2}\frac{\tau}{\sigma}\frac{m\xi_0}{\varepsilon}=\frac{\lambda_C}{\sigma}\frac{\xi_0^2}{\chi_0}\Psi
\end{equation}
times the leading-order contribution, with $\Psi=\omega_0\tau$. Since this is a first-order correction, one would have expected a scaling as $|\bm{\rho}_{\perp}|/\sigma\propto\sqrt{\Psi}$. However, terms proportional to $\bm{\rho}_{\perp}$ in the pre-exponent vanish after integrating over $\bm{\rho}_{\perp}$ and only terms even in $\bm{\rho}_{\perp}$ give a non-vanishing contribution, which explains the scaling as $\Psi$.  Also, the parameter $\theta$ depends on the transverse variation length scale $\sigma$ of the laser beam, such that in the plane-wave case ($\sigma\to\infty$) the parameter $\theta$ vanishes and TFL effects cannot even be estimated starting from a plane-wave model. Finally we note that the fact that, after restoring cgs units, the parameter $\theta$ does not contain $\hbar$ [see Eq. (\ref{theta})] should not confuse, as in any case the correction contains an additional factor $\omega/\varepsilon$, which vanishes for $\hbar\to 0$ [see Eq. (\ref{dW_domega_1})].

The odd number of powers of external field in the pre-exponential function in Eq. (\ref{dW_domega_1}) suggests that a suppression of the first-order correction is expected especially for long pulses, unless specially shaped fields can be employed. For this reason but also in order to confirm $\theta$ as the parameter controlling the TFL interference effects at the lowest orders, we carry out a second-order expansion of the phase $\Phi$ in $\delta_{\perp}$ and perform the analogous analysis as that below Eq. (\ref{Phi_1}) to determine the vector $\bm{R}_{\perp}$. The final result for the phase $\Phi$ up to the second order in $\delta_{\perp}$ is
\begin{equation}
\label{Phi_2}
\begin{split}
&\Phi^{(2)}=\frac{T_-}{2}\bigg\{\frac{m^2\omega}{\varepsilon\varepsilon'}+\frac{1}{\varepsilon}(\bm{I}^{(-)\,2}_{\text{in}}-J^{(-)}_{\text{in}})-\frac{1}{\varepsilon'}(\bm{I}^{(-)\,2}_{\text{out}}-J^{(-)}_{\text{out}})-\frac{\varepsilon}{T^2_-}\bm{\rho}_{\perp}^2-\frac{1}{T_-}\bm{\rho}_{\perp}\cdot\delta_{\perp}\bm{I}^{(+)}_{\text{in}}\\
&\quad-\frac{1}{4\varepsilon}(\delta_{\perp}\bm{I}^{(+)}_{\text{in}})^2-\frac{1}{4T_-}\bm{\rho}_{\perp}\cdot\delta^2_{\perp}\bm{I}^{(-)}_{\text{in}}\\
&\quad-\frac{T_-^2}{4\varepsilon}\left[\frac{1}{\varepsilon}\bigg(I^{(-)}_{\text{in},j}\bm{\nabla}_{\perp}I^{(+)}_{\text{in},j}-\frac{\bm{\nabla}_{\perp}}{2}J^{(+)}_{\text{in}}\bigg)-\frac{1}{\varepsilon'}\bigg(I^{(-)}_{\text{out},j}\bm{\nabla}_{\perp}I^{(+)}_{\text{out},j}-\frac{\bm{\nabla}_{\perp}}{2}J^{(+)}_{\text{out}}\bigg)\right]^2\\
&\quad+\frac{1}{4\varepsilon}\bigg[(\delta_{\perp}\bm{I}^{(+)}_{\text{in}})^2+\bm{I}^{(-)}_{\text{in}}\cdot\delta^2_{\perp}\bm{I}^{(-)}_{\text{in}}-\frac{\delta^2_{\perp}}{2}J^{(-)}_{\text{in}}\bigg]-\frac{1}{4\varepsilon'}\bigg[(\delta_{\perp}\bm{I}^{(+)}_{\text{out}})^2+\bm{I}^{(-)}_{\text{out}}\cdot\delta_{\perp}\bm{I}^{(-)}_{\text{out}}-\frac{\delta^2_{\perp}}{2}J^{(-)}_{\text{out}}\bigg]\bigg\}.
\end{split}
\end{equation}
Looking at $\Phi^{(2)}$ as a function of $\bm{\rho}_{\perp}$, we conclude that the term in the third line is the second-order correction to the first three constant terms in the first line, whereas the terms in the fourth line are the corrections to the zero-order term $-(\varepsilon/2T_-)\bm{\rho}_{\perp}^2$ and then to the TFL. Since the largest corrections arise from the terms containing the quantities $J^{(\pm)}_{\text{in/out}}$ (this is why we have ignored in the discussion the last two small second-order terms in the second line), we see that indeed they are in all cases about $\theta^2$ times the corresponding zero-order terms, as expected from $\theta$ being the controlling parameter. Two short remarks are in order: 1) if $\bm{p}_{\perp}=\bm{0}$, then $J^{(-)}_{\text{in}}=J^{(-)}_{\text{out}}$ and a compensation takes place such that the correction to the TFL acquires an additional factor $\omega/\varepsilon'$; 2) limiting to higher-order even corrections in $\delta_{\perp}$, the $(2n)$th-order correction is $\delta^2_{\perp}$ times smaller than the $(2(n-1))$th-order correction. As we have already pointed out, the reason to compute the second-order expansion of $\Phi$ was to confirm $\theta$ as scaling parameter for the lowest-order corrections, which will be employed below for numerical estimations. However, a self-consistent computation of the second-order expansion of the emission spectrum would require additional terms in the phase arising from the expansion of the states up to the second order in the inverse of the energies, which goes beyond the present semi-quantitative analysis.

The above results indicate that if we consider a feasible setup of an optical ($\lambda_0=1\;\text{$\mu$m}$) FF pulse of peak intensity $I_0=3\times 10^{18}\;\text{W/cm$^2$}$ ($\xi_0\approx 1$), spot radius $\sigma=2\;\text{$\mu$m}$, and of pulse duration $\tau=100\;\text{ps}$ ($\Psi\sim 1.9\times 10^5$ and total pulse energy of about $40\;\text{J}$), colliding with an electron of energy $\varepsilon=8\;\text{GeV}$ ($\chi_0\approx 0.08$), we obtain that $\theta\approx 0.5$ and thus we expect corrections to the leading-order differential probability $dP_0/d\omega$ of the order of $50\,\%$ of its value [see the discussion below Eq. (\ref{theta})]. Although a perturbative approach may be questionable for corrections of the order of $50\,\%$, the above considerations aim to show that for feasible laser and electron parameters we expect the TFL effects to substantially alter the photon differential emission probability, i.e., that values of the parameter $\theta$ of the order of unity are feasible. Indeed, FF pulses with intensities of the order of $10^{14}\;\text{W/cm$^2$}$ have been produced \cite{Turnbull_2018} and intensities beyond the relativistic threshold $\xi_0=1$ are already envisaged \cite{Palastro_2020}. In this respect, we observe that even though the focus moves along the same direction of the electron, the value of the quantum parameter $\chi$ is not suppressed as in the case, e.g., of a plane wave copropagating with an ultrarelativistic electron. The reason is that a FF beam, as that discussed above, has to be thought as being made of waves counterpropagating with respect to the electron but all focused in different points at different times. This can be explicitly verified via the expression of the field given in the Appendix A and recalling that the local value $\chi(x)$ of the quantum nonlinearity parameter is approximately given by $\chi(x)\approx (p_+/m)(|\partial\bm{A}_{\perp}(x)/\partial T|/E_{cr})$ \cite{Di_Piazza_2015}.

In the estimation above it has been implicitly assumed that the electron stays for the whole pulse duration inside the focus. This would be the case if $|\bm{r}_{\perp}|<\sigma$, i.e., for incoming transverse momenta such that $2\theta|\bm{p}_{\perp}|<m\xi_0$. Experimentally, this implies to use a sufficiently collimated electron beam in order to maximize the effects of the TFL. Also, due to the length of the pulses under considerations, we expect a large number of emissions per electron, whereas here the emission of a single photon has been investigated. On the one hand, this does not impact the importance of the present results. Referring to the above example, in fact, due to the relatively small value of $\chi_0$, recoil effects are expected to be moderate such that the number of emitted photons is approximately distributed according to a Poisson distribution with the quantity $P=\int_0^{\varepsilon}d\omega d^4x_+\, dW(x_+)/d\omega$ representing the average number of photons emitted \cite{Glauber_1951}. On the other hand, however, the large number of emissions may significantly increase the angular opening of the electrons and let them exit the FF beam laterally. In the range of parameters at hand the number $n_{\gamma}$ of photons emitted can be estimated as $\alpha n_L$ \cite{Ritus_1985,Baier_b_1998,Di_Piazza_2012}. In the above example, about 200 photons are emitted. Hence, being the photons randomly emitted within a cone of angular aperture $m/\varepsilon$, the actual condition on $\bm{p}_{\perp}$ has to be about $\sqrt{n_{\gamma}}\approx 15$ more restrictive than the above one.

\section{Conclusions}
In conclusion, we have studied nonlinear Compton scattering by an ultrarelativistic electron counterpropagating with respect to a flying focus laser beam in the regime $\xi_0\sim 1$ and $\chi_0\lesssim 1$ and we have shown that the effects of the transverse formation length of radiation are potentially amplified by orders of magnitudes as compared to a fixed focus beam. After identifying the parameter controlling these effects at the lowest orders, we have shown that values of this parameter of the order of unity can be reached and then substantial corrections to the emission probability are expected for tightly-focused optical flying focus fields of peak intensity $\sim 10^{18}\;\text{W/cm$^2$}$ and duration $\sim 100\;\text{ps}$, with the focus counterpropagating at the speed of light with respect to the laser beam. Finally, we point out that the unique structure of flying focus beams offers further potential applications in strong-field classical and quantum electrodynamics, providing a new experimental tool to test these theories in the high-intensity regime.

\begin{acknowledgments}
The author gratefully acknowledges helpful discussions with D. H. Froula, K. Z. Hatsagortsyan, J. P. Palastro, F. Qu\'{e}r\'{e}, Y. I. Salamin, and G. Sarri. This publication was also supported by the Collaborative Research Centre 1225 funded by Deutsche Forschungsgemeinschaft (DFG, German Research Foundation) - Project-ID 273811115 - SFB 1225.
\end{acknowledgments}

\appendix

\section{An exact solution of Maxwell's equation suitable for describing a flying focus field with the focus counterpropagating at the speed of light with respect to the pulse}
Here, we report an explicit expression of the electromagnetic field, which is an exact solution of Maxwell's equations and which may be used to describe the main features of a flying focus (FF) field, with the focus moving at the speed of light in the opposite direction as that of the pulse propagation [see Ref. \cite{Palastro_2018} for a more accurate expression of the electromagnetic field of a FF beam]. We recall that we consider a laser beam whose focal plane corresponds to the $x\text{-}y$ plane and whose wave vector at the center of the focal area points along the negative $z$ direction. Thus, the focus moves at the speed of light along the positive $z$ direction. We closely follow the approach reported in Ref. \cite{Esarey_1995} by adapting it to the axial gauge used here in which the four-vector potential $A^{\mu}(x)=0$ of the background field satisfies the Lorenz -gauge condition $\partial_{\mu}A^{\mu}(x)=0$ and the additional constraint $A_-(x)=0$, together with the free wave equation $\partial_{\mu}\partial^{\mu}A^{\nu}(x)=0$ [in Ref. \cite{Esarey_1995} the gauge $A^0(x)=0$ is chosen]. 

First, we write the four-vector potential $A^{\mu}(x)$ in the momentum space corresponding to the light-cone variable $T$, which is convenient since the wave vector at the center of the focal area points along the negative $z$ direction:
\begin{equation}
A^{\mu}(T,\bm{x}_{\perp},\phi)=\int\frac{dk}{2\pi}\tilde{A}^{\mu}(k,\bm{x}_{\perp},\phi)e^{-ikT}+\text{c.c.},
\end{equation}
where $\text{c.c.}$ stands for complex conjugated. Since $A_-(x)=0$, the Lorenz-gauge condition $\partial_TA_+(T,\bm{x}_{\perp},\phi)+\bm{\nabla}_{\perp}\cdot\bm{A}_{\perp}(T,\bm{x}_{\perp},\phi)=0$, allows to determine $\tilde{A}_+(k,\bm{x}_{\perp},\phi)$ as a function of $\tilde{\bm{A}}_{\perp}(k,\bm{x}_{\perp},\phi)$ as
\begin{equation}
\label{A_+}
\tilde{A}_+(k,\bm{x}_{\perp},\phi)=-\frac{i}{k}\bm{\nabla}_{\perp}\cdot\tilde{\bm{A}}_{\perp}(k,\bm{x}_{\perp},\phi),
\end{equation}
such that the only unknown quantity to be determined is $\tilde{\bm{A}}_{\perp}(k,\bm{x}_{\perp},\phi)$. The free wave equation for $\tilde{\bm{A}}_{\perp}(k,\bm{x}_{\perp},\phi)$ reads
\begin{equation}
-2ik\frac{\partial\tilde{\bm{A}}_{\perp}(k,\bm{x}_{\perp},\phi)}{\partial\phi}-\bm{\nabla}_{\perp}^2\tilde{\bm{A}}_{\perp}(k,\bm{x}_{\perp},\phi)=0.
\end{equation}
Having in mind a focused Gaussian beam, we notice that this equation admits the exact analytical solution \cite{Esarey_1995}
\begin{equation}
\tilde{\bm{A}}_{\perp}(k,\bm{x}_{\perp},\phi)=\tilde{\bm{A}}_0f(k)\frac{e^{-\frac{\bm{x}_{\perp}^2}{2\sigma^2(1+i\phi/l_k)}}}{1+i\phi/l_k},
\end{equation}
where $\tilde{\bm{A}}_0$ is a real constant related to the field amplitude, $\sigma$ is a real constant related to the spot radius of the field, $l_k=k\sigma^2$, and $f(k)$ is an arbitrary complex function of $k$ describing the field pulse shape in $T$. By considering a pulse with a Gaussian shape in $T$, we choose
\begin{equation}
f(k)=\frac{k}{k_0}e^{-(k-k_0)^2\tau^2/8},
\end{equation}
where the real constants $k_0$ and $\tau$ will be related with the central angular frequency and pulse length of the field, respectively, and where the prefactor $k/k_0$ has been inserted in such a way that, even if we don't need it here, the corresponding expression of  $\tilde{A}_+(k,\bm{x}_{\perp},\phi)$ is well behaved [see Eq. (\ref{A_+})]. Finally, by indicating as $E_0$ and $\omega_0$ the amplitude and the central angular frequency of the resulting electric field of the pulse, we can appropriately rewrite the exact solution of the wave equation as
\begin{equation}
\label{A}
\bm{A}_{\perp}(T,\bm{x}_{\perp},\phi)=\bm{E}_0\frac{\tau}{\sqrt{2\pi}}\text{Re}\int\frac{d\omega}{\omega_0}\frac{\omega}{\omega_0}e^{-(\omega-\omega_0)^2\frac{\tau^2}{2}-2i\omega T}\frac{e^{-\frac{\bm{x}_{\perp}^2}{2\sigma^2(1+i\phi/l_{\omega})}}}{1+i\phi/l_{\omega}}
\end{equation}
where $\bm{E}_0=\sqrt{\pi/8}\omega_0\tilde{\bm{A}}_0/\tau$, where $l_{\omega}=2\omega\sigma^2$ and where we performed the change of variable $k=2\omega$ ($\omega_0=k_0/2$). 

The integral in Eq. (\ref{A}), although representing an exact solution of Maxwell's equations, cannot be taken analytically and approximated methods have to be employed. If we limit the consideration to long pulses such that $\omega_0\tau\gg 1$, we can approximately write
\begin{equation}
\label{A_approx}
\bm{A}_{\perp}(T,\bm{x}_{\perp},\phi)\approx\bm{E}_0\frac{\tau}{\sqrt{2\pi}}\text{Re}\int\frac{d\omega}{\omega_0}e^{-(\omega-\omega_0)^2\frac{\tau^2}{2}-2i\omega T}\frac{e^{-\frac{\bm{x}_{\perp}^2}{2\sigma^2(1+i\phi/l_{\omega_0})}}}{1+i\phi/l_{\omega_0}}.
\end{equation}
Now the integral in $\omega$ is Gaussian and it can be taken analytically, giving the final result
\begin{equation}
\label{A_appr}
\bm{A}_{\perp}(T,\bm{x}_{\perp},\phi)\approx \frac{\bm{E}_0}{\omega_0}e^{-2T^2/\tau^2}\frac{\sigma}{\sigma_{\phi}}e^{-\bm{x}^2_{\perp}/2\sigma^2_{\phi}}\cos\left[2\omega_0T-\frac{\bm{x}_{\perp}^2}{2\sigma^2_{\phi}}\frac{\phi}{l_{\omega_0}}+\arctan\left(\frac{\phi}{l_{\omega_0}}\right)\right],
\end{equation}
where $\sigma_{\phi}=\sigma\sqrt{1+\phi^2/l_{\omega_0}^2}$. We notice that the different but similar solution found in Ref. \cite{Esarey_1995} was interpreted as describing an ultrashort laser pulse with fixed focus [see also Refs. \cite{Salamin_2015,Li_2016}] such that it was considered to be accurate only for pulse lengths much shorter than the central Rayleigh length \cite{Li_2016}. Here, Eq. (\ref{A_appr}) is interpreted as describing a long FF pulse with the focus moving at the speed of light backwards as compared to the phase velocity, which precisely fits the physical situation of interest in the main text without additional restrictions apart from the long-pulse condition $\omega_0\tau\gg 1$.

%


\end{document}